\documentclass{ws-procs9x6}

\usepackage{ws-toc,ws-multind}
\makeindex{author}

\usepackage{graphicx}
\DeclareGraphicsExtensions{.pdf,.eps,.png,.jpg}
\usepackage[caption=false]{subfig}

\allowdisplaybreaks

\def\be{\begin{eqnarray}}
\def\ee{\end{eqnarray}}

\def\roughly#1{\mathrel{\raise.3ex\hbox{$#1$\kern-.75em%
\lower1ex\hbox{$\sim$}}}}

\def\gsim{\roughly>}

\def\bi{\bibitem}

\begin{document}

\title{Compact Star Matter: EoS with New Scaling Law}

\author{Kyungmin Kim, Hyun Kyu Lee$^*$, Jaehyun Lee}
\aindx{Kim, K.}\aindx{Lee, H. K.}\aindx{Lee, J.}
\address{Department of Physics, Hanyang University, Seoul 04763, Korea\\
\email{$^*\!$E-mail: hyunkyu@hanyang.ac.kr}}

\begin{abstract}

 In this paper we present  a simple discussion on the properties of compact stars  using an EoS obtained in effective field theory anchored on scale and hidden-local symmetric Lagrangian endowed with topology change and  a unequivocal prediction on the deformation of the compact star, that could  be measured in gravitational  waves.   The objective is not to offer a superior or improved EoS for compact stars but to confront with a forthcoming astrophysical observable the given model formulated in what is considered to be consistent with the premise of QCD. The model so obtained is found to satisfactorily describe the observation of a 2-solar mass neutron star \cite{demorest,antoniadis} with a minimum number of parameters.  Specifically the observable we are considering in this paper is the tidal deformability parameter $\lambda$ (equivalently the Love number $k_2$), which affects  gravitational wave forms at the late period of inspiral stage.  The  forthcoming  aLIGO and aVirgo observations of gravitational waves from binary neutron star system will  provide a valuable guidance for arriving at a better understanding of highly compressed baryonic matter.

\end{abstract}




\section{INTRODUCTION}

The observation of 2-solar mass neutron stars\cite{demorest,antoniadis} seems
to indicate that the equation of state(EoS) for compact stars
needs to be sufficiently stiffer to accommodate the mass larger
than 1.5-solar mass. Moreover it requires the detailed information on the hadronic
matter   at higher density than the normal nuclear density, $n_0$, which seems to be however much
higher beyond the reach of presently planned terrestrial laboratories.  After the recent detections of gravitational waves from binary black holes\cite{PRL}, the expectation of detecting  gravitational waves from a binary neutron stars and/or a  black hole-neutron star binary  becomes very optimistic than ever.  The gravitational waves emitted during the binary inspiral phase up to merging can provide us the information on the dense hadronic matter of higher density at the core of compact stars.

The nuclei with large atomic numbers are  already obvious examples of  highly dense matters composed of nucleons, $n \sim n_0 = 0.16/\textrm{fm}^{-3}$.  The effective theories of nucleons   have been developed and constrained by experimental data available up to and slightly above the normal nuclear density, $n_0$. Hence they are
fairly well controlled theoretically and experimentally.  but the high density regime much above $n_0$ is more or less uncharted both experimentally and theoretically.  Compact objects such as neutron stars are supposed to have higher core density
than the normal nuclear density, $n_0$.  Roughly for neutron star with mass $\sim 1.5 M_{\odot}$ the relevant density at the center is likely  around  $2 n_0-3 n_0$  and for the mass $\sim 2 M_{\odot}$ the density is supposed to be larger than $\sim 5n_0$.  On top of the possibilities of getting high density nuclear matter at terrestrial laboratories, for example FRIB(USA), FAIR(Germany), J-Parc(Japan) and RAON(Korea) in near future, the possible detections of gravitational waves from binary neutron stars(or binary neutron star black hole) are believed to be  promising probes of the high density interior of neutron stars.  Theoretically the success of low density effective theories univocally up to the normal nuclear density  seems not to  guarantee the similar success at higher density hadronic matter since the predictions of mass, radius, symmetry energy and  deformability to name a few  diverse from each other drastically beyond the normal nuclear density.  In this sense we are now at the very exciting period of foreseeing the opportunity of constraining  theories at higher density by experiments and observations.

For the highly dense hadronic matter, recently we proposed a new approach\cite{Dong,LPR,Paeng} in which we   try to formulate a field theory framework wherein both low and high density regimes are treated on the same footing. To cover both regimes in a consistent way in a unified field theoretic approach seems like a tall order.
In this work, we discuss the physical properties of stellar matter using the newly proposed  scheme of  a new scaling law(BLPR) in medium\cite{byp,lr} all the way from normal nuclear density up to the higher density at the core of a neutron star. With a confrontation with the observed massive stars\cite{Dong}, we discuss the physical properties of stellar matter with the EoS obtained therein.  The relevant quantities are mass, radius and  deformability parameters,  which could be  constrained  by the  gravitational wave forms  emitted during the binary inspiraling  phase. The aim here is then to confirm or falsify the strategies taken and assumptions made in Dong {\it et al.}\cite{Dong} and Lee {\it et al.}\cite{LPR} and also to find the directions  to be taken in constructing the correct effective theory at higher density.

In section II,  the basic concept of unified approach in this work is discussed.  Using the minimal effective lagrangian in the frame work of relativistic mean field,  the equation of state of compact star with neutrons, protons, electrons and muon in weak equilibrium and charge neutrality condition is discussed in section III and the mass and radius are estimated. Tidal deformation with new stiffer EoS is discussed with the observational possibility in gravitational wave detections at aLIGO\cite{aligo} and aVirgo\cite{avirgo} in section IV. The results are summarized  in section V. We use units in which $c=G=1$ and the notation in which Minkowski metric $\eta_{\mu\nu} = \mathrm{diag}[-1,\ 1,\ 1,\ 1]$.

\section{Unified approach with  new scaling}

The unified field theoretic approach is formulated using an effective theory which has sufficient  number of  degrees of freedom and parameters to be able to implement desired symmetries  in high density regime or at the   critical density, $n_c > n_0$.  The first step is to construct the functional forms of  parameters of the effective theory  in terms of the variables in QCD, for example, quark and gluon  condensates.   by matching procedures near at  QCD scale .  Then, if we can compute  the density dependence of QCD condensates, they   automatically determine the density dependence of parameters in the effective theory.  In addition the vacuum can be  characterized by the expectation value of the relevant field, in this work scalar field $<\chi>$, which is supposed to depend on the density.  These are the  intrinsic density dependencies(IDD) inherited from the quark/gluon  condensates and scalar condensate.  At low density (up to nuclear matter density) the success of chiral perturbation theory(ChPT) and low energy theorems implies  that the QCD matching is implicitly taken care already.  But at higher density,  QCD matching  or IDD  will play a very crucial role since the naive extrapolation of ChPT, in which the QCD matching  is implicit and  therefore hidden,  may go anywhere  if not guided by the explicit QCD matching constraints.

The  first step forward to that goal has been made using the effective lagrangian with minimum number of degree of freedom, which are  the pseudo scalar mesons $\pi$, and scalar meson $\sigma$, vector mesons $\rho$ and nucleon $N$. We adopt the relativistic mean field(RMF) approach, which are found to work remarkably well for finite and nuclear matter as well as  for highly dense matter inside compact stars. The  success of RMF can be understood that RMF captures the physics of the Landau Fermi liquid fixed point  at  which  nuclear matter is located.
It is assumed that the symmetries hidden at lower density on top of underlying chiral symmetry are the hidden local symmetry and the scale symmetry, which  are supposed to  be manifested explicitly eventually at the critical density higher than normal nuclear density.   The details are  reviewed in recent  articles  \cite{LPR,Paeng,rho}.

When BR scaling(old-BR)\cite{BR} is applied to the neutron-star calculation using  realistic NN potentials\cite{tom}, the mass is estimated to be in the range, $1.2M_{\odot} \sim 1.8 M_{\odot}$, which is apparently   less than 2$M_{\odot}$.  New BR/BLPR scaling has been proposed\cite{Dong,RhoG} to incorporate the change in  topology of  the crystal structure of skyrmions, skyrmion $\rightarrow$ half-skyrmion\cite{byp}. Suppose the threshold density, $n_{1/2}$,  is higher but not so higher than the normal nuclear density, $n_0$, then we expect the physical effect  of such topology change on nuclear matter at the density $n \geq n_{1/2}$. More elaborated effective lagrangian including $\omega$-meson has been discussed by Paeng {\it et al.}\cite{Paeng} but  in this work we use  the simpler version discussed by Dong {\it et al.}\cite{Dong}.

 The drastic change in the symmetry energy observed at the density $n_{1/2}$ \cite{byp,lr}  can be translated into the parameter changes of the Lagrangian, a new scaling. With new scaling,  new-BR/BLPR scaling, one can understand the origin of the drastic change in symmetry energy is due to the substantial change in tensor force, disappearance of $\rho$  tensor component at higher density.  The new scaling is incorporated into the $V_{lowk}$ - implemented EFT approach to calculate the equation of state and the mass - radius relation of a compact object of a pure  neutron matter.  The stiffer equation of state has been obtained as expected and the mass can be as large as $2.4 M_{\odot}$\cite{Dong}, which seems to be consistent with recently observed high mass neutron stars.  In this work we take a more realistic approach for the compact star with electrons, protons and neutrons,  which are believed to be in weak equilibrium, rather than pure neutron matter. Near the surface of star, which is supposed to be  in lower density region, $n < 0.5 \, n_0$, we adopt the equation of state  used in  Hebeler {\it et al.}\cite{Astro.J/773/11}.

It is  assumed, in the range of density we are considering, the energy density of asymmetric nuclear matter($n_p \neq n_n$ or $n_p/n \neq 1/2$) can be described by the conventional form in terms of symmetry energy, $S(n)$, as given by

\be
\epsilon_{nuc} (n,x) &=&  \epsilon_{nuc} (n,x=1/2) + n(1-2x)^2 S (n)  \label{enuc}
\ee
where  $x \equiv n_p /n$ is the fraction of proton density.
Then the symmetry energy factor $S(n)$ can be obtained  by
\be
\label{sym_e_fac}
S(n) &=& \epsilon_{nuc} (n,0)/n -  \epsilon_{nuc} (n,x=1/2)/n
\ee
which is equivalent to the difference of binding energy per nucleon between  the symmetric nuclear matter ($x=1/2$) and  the  neutron matter ($x=0$).  The pressure of nuclear matter is given by $p_{nuc} = n^2\frac{\partial \epsilon(n)/n}{\partial n}$.
In this work, we use the corresponding  binding energy   and the symmetry energy factor obtained by Dong {\it et al.} \cite{Dong} with the new scaling. In the  weak equilibrium, the proton fraction, $x$,  is determined  essentially by  the chemical potential difference between proton and neutron ,
\be
\mu_n-\mu_p = 4(1-x)S(n) \label{munp}
\ee
together with charge neutrality condition, $n_p = n_e + n_\mu$.

Now given EoS for energy density, $\epsilon$, and pressure, $p$, the radius, $R$, and mass, $m(R)$, can be determined by integrating the Tolmann-Oppenheimer-Volkoff (TOV) equations \cite{PhysRev/55/364, PhysRev/55/374}.  The equations are integrated up to the radius of the star, $R$, where $p(R) = 0$, and the mass of the star is determined by $m(R)$.  The masses and radii, which depend on the equation of state,  are important  in predicting the gravitational waves emitted  from the coalescing binary neutron stars. During the  inspiral period of binary neutron stars,  tidal distortions of neutron stars are expected and the resulting gravitational wave is expected to carry the corresponding information of equation of states\cite{PRD/81/123016}.

The tidal deformability of polytropic EoS, $p = K \epsilon^{1+1/n}$, where $K$ is a pressure constant and $n$ is the polytropic index, were evaluated by Flanagan and Hinderer\cite{PRD/77/021502, ApJ/677/1216} and by others in more detail\cite{PRD/80/084035, PRD/80/084018}. Recent works by Reads {\it et al.}\cite{read} and Hotokezaka {\it et al.}\cite{HKSS} demonstrate the measurability of the tidal deformation to constrain the EoS.   In this work, we calculate the mass-radius relation  and the tidal deformability using the stiffer EoS, which has been recently proposed with  new scaling law(BLPR)\cite{Dong,Paeng}.


\section{Compact star composed of  n, p, e, and $\mu$}

The asymmetry of neutron and proton numbers at high density, dictated by the chemical potential difference,    inevitably leads to the weak equilibrium configuration with electron and muon with neutrinos escaped. It can be summarized by the  relation between  chemical potentials given by
\be
\mu_n-\mu_p = \mu_e = \mu_{\mu}. \label{munpe}
\ee
The chemical potential difference between neutron and proton should be the same as the electron chemical potential. The last equality is due to the muon emergence at higher density when  the chemical potential difference from neutron and proton becomes larger than the muon mass.
With  the charge carriers,  proton, electron and muon, the local charge neutrality condition is given by
\begin{equation}
\label{neutral}
n_p = n_e + n_\mu ~.
\end{equation}
Using Eq.\ (\ref{munp}), Eq.\ (\ref{munpe}), and Eq.\ (\ref{neutral}) can be solved to get the carrier densities at a given density.  The density dependence of proton fraction, $x$, is shown in Fig. 1(a). One can see that the  proton fraction increases significantly as density increases.

The total energy density and pressure  are given by
\be
\epsilon (n,x) &=& \epsilon_{nuc} + \epsilon_{lep} ~, \\
p (n,x) &=& p_{nuc} + p_{lep} ~.
\ee
The energy density, $\epsilon_{lep}$,  and the pressure, $p_{lep}$  are given by the degenerate  fermi gas of  the leptons(electron and muon) assuming a cold compact star ($T\sim0$).

\begin{figure}[!t]
\subfloat[]{\label{fig3-1}\includegraphics[width=2.25in]{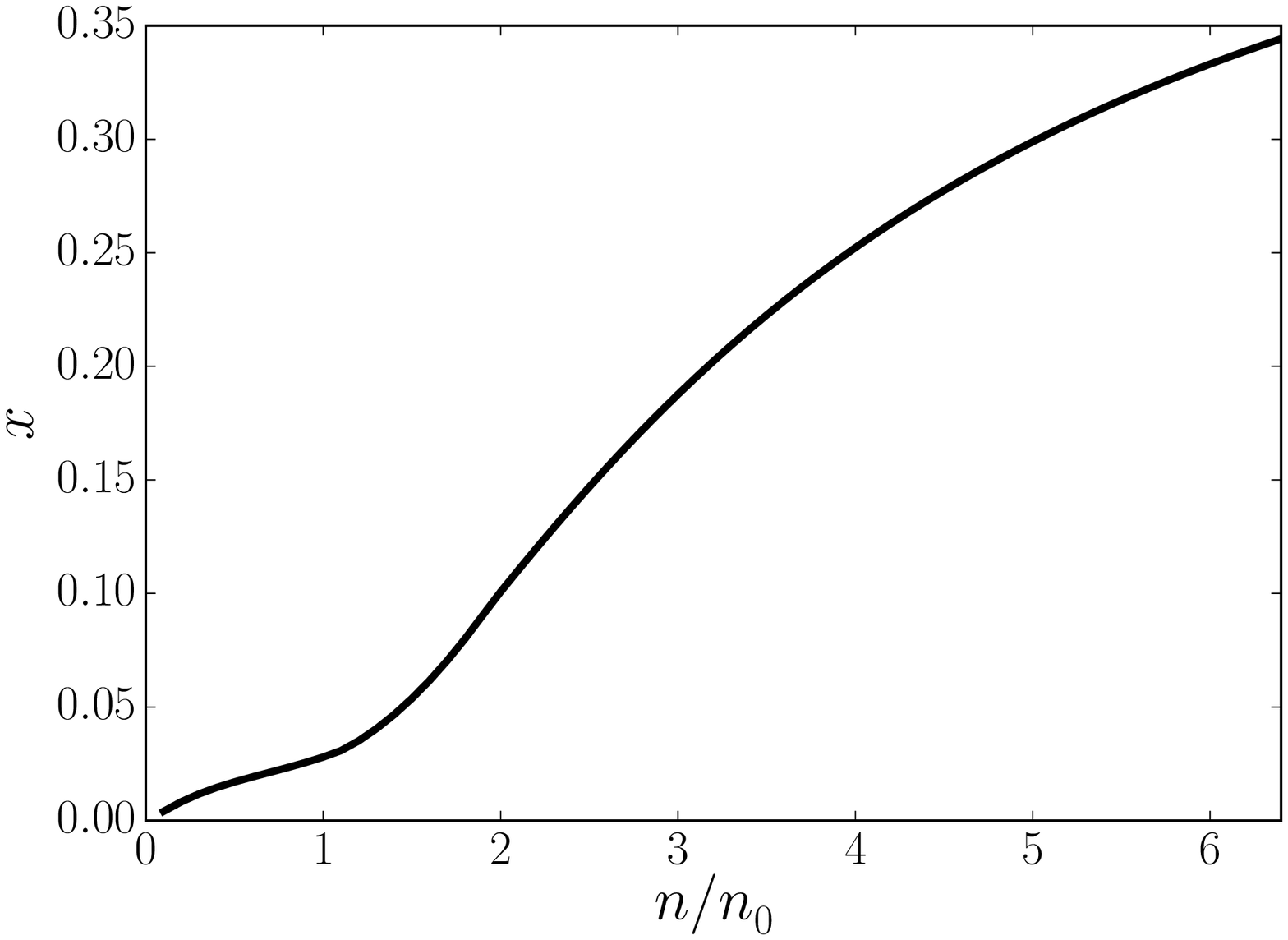}}
\subfloat[]{\label{fig3-2} \includegraphics[width=2.25in]{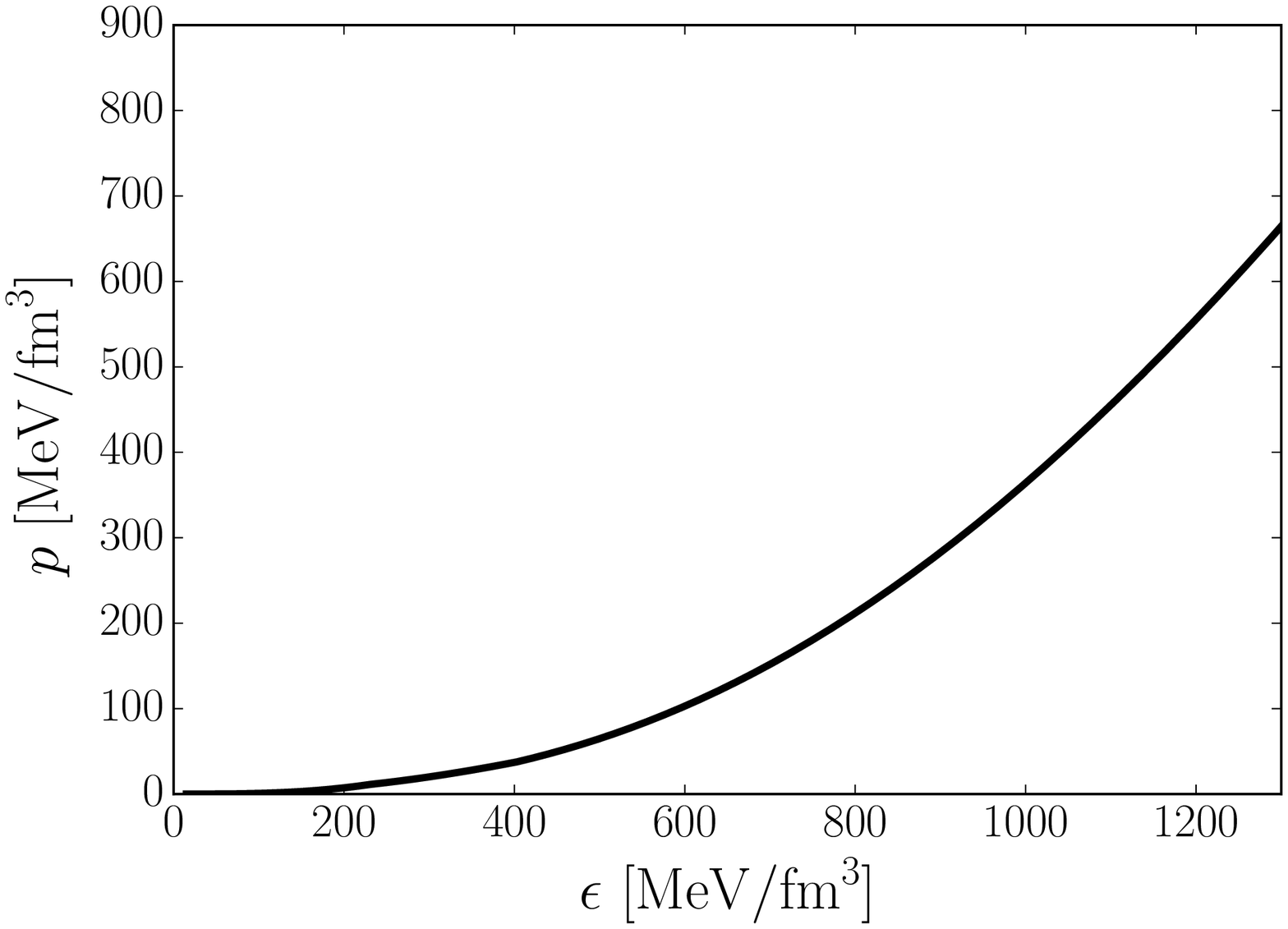}}
\caption{(a) The proton ratio $x$ as a function of nuclear density $n$.
(b) The relation between energy density and pressure for an $np$ asymmetric configuration.
}
\label{fig3}
\end{figure}

The resulting equation of state is shown as a pressure-energy density diagram in Fig. \ref{fig3}.
The causality limit, $c_s \leq c$,  constrains   the highest density, $n_c$, beyond which  the stiffer EoS used is no more valid. For EoS used in  this work,  it  is found to be $n_c \sim 5.7\ n_0$.

For a static and spherically symmetric astrophysical compact star, the metric is given by
\be
ds^2 &=& -e^{\Phi(r)}dt^2 + e^{\Lambda(r)} dr^2 + r^2 d\theta^2 + r^2\sin^2 \theta d\phi^2. \nonumber \\ \label{smetric}
\ee
where $\Lambda$ can be expressed in terms of a radial-dependent  mass parameter, $m(r)$, introduced by
\be
 e^{\Lambda(r)}  = (1-\frac{2m}{r})^{-1}.
 \ee
Assuming a perfect-fluid stellar matter,
\be
T_{\mu\nu} = (\epsilon + p) u_\mu u_\nu + p g_{\mu\nu},\label{perfect}
\ee
the relativistic hydrodynamic equilibrium is governed  by the TOV equation
\be
\frac{dm}{dr} &=& 4\pi r^2 \epsilon \label{TOV1}\\
\frac{dp}{dr} &=& - ( \epsilon + p ) \frac{ m + 4\pi r^3 p}{r ( r - 2 m )}\label{TOV2} \\
\frac{d\Phi}{dr} &=& - \frac{1}{\epsilon +p} \frac{dp}{dr}.\label{TOV3}
\ee
where $\epsilon$ and $p$ are energy density and pressure at $r$ respectively and  $u^\mu = d x^\mu /d \tau$ is the fluid's four-velocity. $m(r)$ is the mass enclosed inside the radius $r$. We can calculate the mass of compact star, $M$,  and its radius, $R$,  by integrating the TOV equation up to $p(R)=0$ and we get the profile of $\Phi(r), m(r)$ and $p(r)$.

\begin{figure}[!t]
\subfloat[]{\label{fig2a} \includegraphics[width=2.25in]{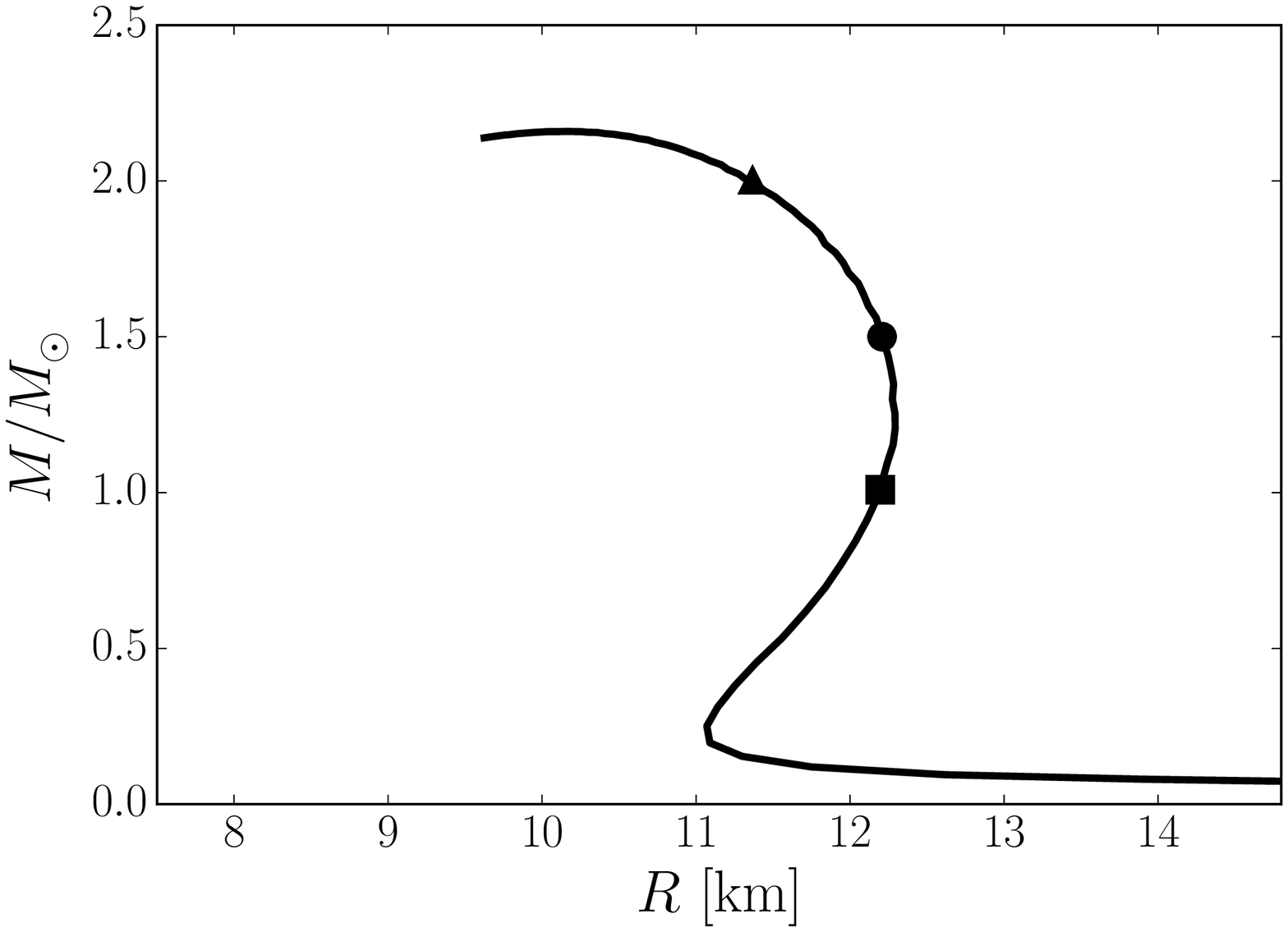}}
\subfloat[]{\label{fig2b} \includegraphics[width=2.25in]{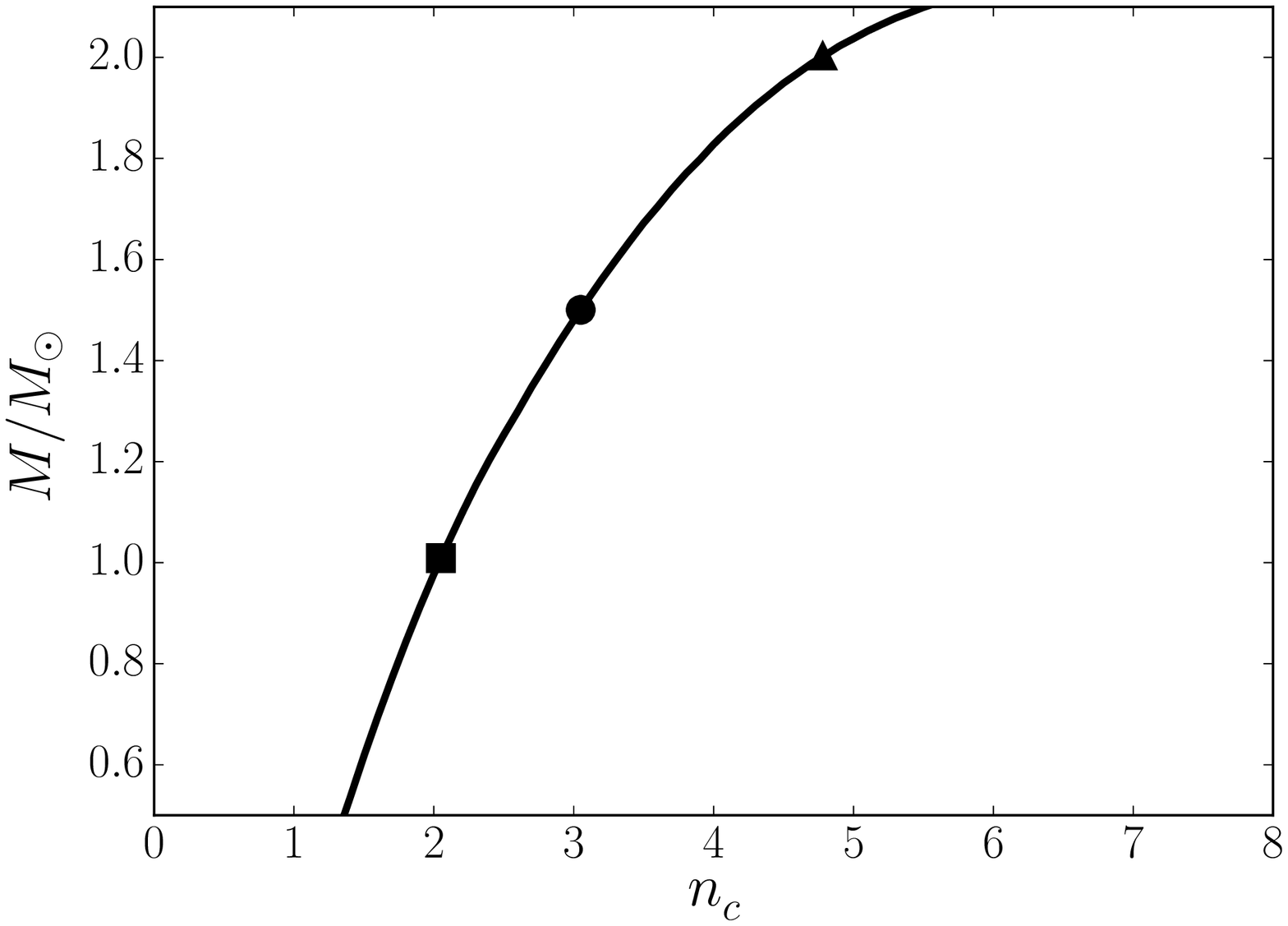}}
\caption{(a) Mass($M$)-Radius($R$) curve. The filled-square, filled-circle, and filled-triangle correspond to  $M=1.0 M_\odot$ , $M=1.5 M_\odot$ , and $M=2.0 M_\odot$ respectively. (b) $n_c$ vs. Mass.}
\label{fig5}
\end{figure}

The EoS of $np$ asymmetric configuration is used to solve the TOV equation resulting in the mass-radius curve shown in Fig.\ \ref{fig5}.  For $np$ asymmetric configuration, the possible maximum mass is estimated  to be $M \sim 2.1 M_\odot$ with the radius $R \sim 11$km, where the central density is about $5.7 \, n_0$. For pure neutron matter~\cite{Dong}, the possible maximum mass is approximately $M \sim 2.4 M_\odot$ with the radius $R \sim 12$km and $n \sim 4.7 n_0$. In Fig.\ \ref{fig2a},  the filled-square, filled-circle and filled-triangle correspond to  $M=1.0 M_\odot$ , $M=1.5 M_\odot$ , and $M=2.0 M_\odot$, respectively. The compactness $C=\frac{M}{R}$ in the range of mass $1.0 - 2 M_{\odot}$ is found to be 0.12 - 0.26 and 0.14 for $1.4 M_\odot$
Fig.\ \ref{fig2b} shows the relation between central density $n_c$ and the radius and between $n_c$ and the mass of the star.

\section{Tidal Deformability}
When a compact star is placed in a static external field, we suppose a star in a  spherically symmetric configuration is then deformed by the external   field.  The asymptotic expansion of the  metric at large distances $r$ from the star defines the quadrupole moment, $Q_{ij}$,  and the external tidal field, $\mathcal{E}_{ij}$, as expansion coefficients \cite{PRD/58/124031} given by
\be
-\frac{1 + g_{00}}{2} &= &-\bigg[ \frac{m}{r} + \frac{3}{2} \frac{Q_{ij}}{r^3} n^i n^j + \cdots \bigg] \nonumber \\
 & &+ \frac{1}{2} \mathcal{E}_{ij} r^2 n^i n^j + \cdots ~, \label{quad}
\ee
where $n^i = x^i / r$ and $Q_{ij}$ and $\mathcal{E}_{ij}$ are both symmetric and traceless \cite{PRD/58/124031}.

The deformability parameter $\lambda$ is defined by
\begin{equation}
\label{def}
Q_{ij} = - \lambda \mathcal{E}_{ij} ~,
\end{equation}
which depends on the EoS of the nuclear matter and provides the information  how easily the star is deformed.
The deformability parameter can be reexpressed by the dimensionless Love number, $k_2$, defined by
\be
\lambda = \frac{2 k_2}{3}R^5 \label{k2}.
\ee

In general,  the  linearized perturbations of the metric caused by an external field is given by \cite{ApJ/149/591},
\begin{equation}
g_{\mu\nu} = g_{\mu\nu}^{(0)} + h_{\mu\nu} ~,\label{hmunu}
\end{equation}
where $g_{\mu\nu}^{(0)}$ is a unperturbed metric,
\be
g_{\mu\nu}^{(0)} &=& \mathrm{diag} \big[- e^{2 \Phi (r)},\ e^{2 \Lambda (r)},\ r^2,\ r^2 \sin^2{\theta} \big] \label{gmunu0}
\ee
where $\Phi(r)$ and  $\Lambda(r)$ will be determined by the stress-energy tensor configuration discussed below.
$h_{\mu\nu}$ is a linearized perturbation, which carries  the information of $Q_{ij}$ and $\mathcal{E}_{ij}$ in Eq.\ (\ref{quad}). Since we will be considering  the early stage of binary inspiral before the merging stage, the leading order tidal effects with even parity, $l=2$, is dominated \cite{PhysRev/108/1063}.  In the Regge-Wheeler gauge, the static and  even-parity perturbation  with $l=2$ denoted by $H(r)$ and $K(r)$ can be written in the following form \cite{ApJ/677/1216}
\be
h_{\mu\nu} &=& \mathrm{diag} \big[ -e^{2 \Phi (r)} H(r) Y_{20} (\theta, \phi), \nonumber \\
&&\qquad \quad -e^{2 \Lambda (r)} H(r) Y_{20} (\theta, \phi), \nonumber \\
&&\qquad \quad -r^2 K(r) Y_{20} (\theta, \phi), \nonumber \\
&&\qquad \quad -r^2 \sin^2{\theta} K(r) Y_{20} (\theta, \phi) \big] ~. \label{hmunu}
\ee
$K(r)$ is related to $H(r)$,
\be
K'(r) = H'(r) + 2H(r) \Phi'(r),
\ee
where the prime $'$ denotes the differentiation $d / dr$.

On the other hand, the non-vanishing component of the perturbation of stress-energy tensor, $\delta T_{\mu\nu}$, due to the tidal deformation, corresponding to $l=2, m=0$ metric perturbation are given by
\be
\delta T_0^0 = - \delta \epsilon(r) Y_{20},  ~~~ \delta T_i^i = \delta p(r) Y_{20}  ~.\label{deltaT}
\ee
Using  the linearized Einstein equations,
\begin{equation}
\delta G_{\mu\nu} = 8 \pi \delta T_{\mu\nu} ~,
\end{equation}
where $G_{\mu\nu}$ is Einstein tensor, we  obtain the differential equation for $H(r)$:
\be
&H''& + \left( \frac{2}{r} + \Phi' - \Lambda' \right) H'  \nonumber \\
&+& \bigg \{ 2 ( \Phi'' - \Phi'^2 ) - \frac{6}{r^2} e^{2 \Lambda} + \frac{3}{r} \Lambda'  \nonumber \\
&+& \frac{7}{r} \Phi' - 2 \Phi' \Lambda' + \frac{f}{r} ( \Phi' + \Lambda' ) \bigg \} H  = 0 ~, \label{ODE}
\ee
where we introduce $f(r)$ for $\delta T_{\mu\nu}$,  given by
\be
 f(r) = \frac{d\epsilon}{dp}. \label{deltae}
\ee

At the asymptotic distance from the stellar matter,  $r \gg M $, where  $T_{\mu\nu} = 0$, $g_{00}$    can be approximated
\be
-\frac{1 + g_{00}}{2} \rightarrow -\bigg[ \frac{M}{r} - \frac{1}{2} H^{asympt}(r)Y_{20}\bigg] , \label{Hasympt}
\ee
where  $ H^{asympt}(r)$ is the solution of Eq.\ (\ref{ODE}) at asymptotic distance given by
\be
H^{asympt} &=& \frac{8}{5} \left( \frac{M}{r} \right)^3 c_1 + \left( \frac{r}{M} \right)^2 c_2 ~ \cdots ~.\label{asympsol}
\ee
One can determine $c_1$ and $c_2$ using the continuity of $H(r)$ and $H'(r)$ at the boundary, $r=R$,  both for interior and outside solutions of Eq.\ (\ref{ODE}).
Then we can read out the deformability parameter by comparing  Eq.\ (\ref{asympsol}) and  Eq.\ (\ref{quad}),
\be
\lambda  = \frac{8}{15} M^5 \frac{c_1}{c_2}. \label{c1c2}
\ee
Using the continuity condition the  $l=2$ deformability $\lambda$ can be written explicitly\cite{ApJ/677/1216} in terms of the compactness $C=M/R$ and $y= R H'(R) / H(R)$,
\be
\lambda_2 &=& \frac{16}{15} R^5 C^5 (1 - 2C)^2 [2 + 2C(y - 1) - y] \nonumber \\
&& \times \Big\{ 2C [6 - 3y + 3C(5y - 8)] \nonumber \\
&& + 4C^3[13 - 11y + C(3y - 2) + 2C^2(1 + y)] \nonumber \\
&& + 3(1 - 2C)^2 [2 - y + 2C(y - 1)] \nonumber \\
&& \qquad \qquad \qquad \qquad \qquad \times \ln{(1-2C)} \Big\}^{-1}. \label{lovek2}
\ee

\begin{figure}[!t]
\subfloat[]{\label{fig6-3} \includegraphics[width=2.25in]{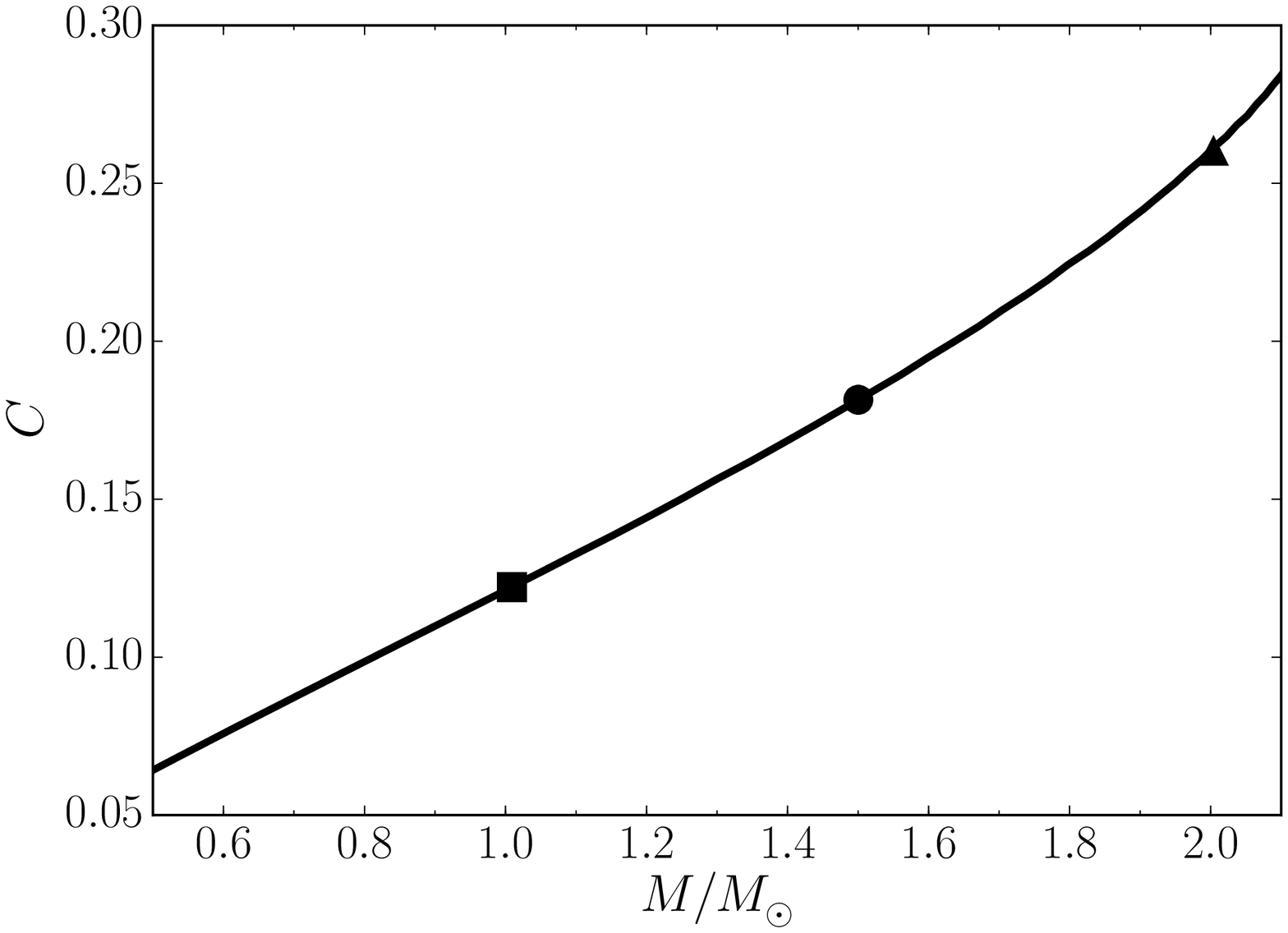}}
\subfloat[]{\label{fig7-3} \includegraphics[width=2.25in]{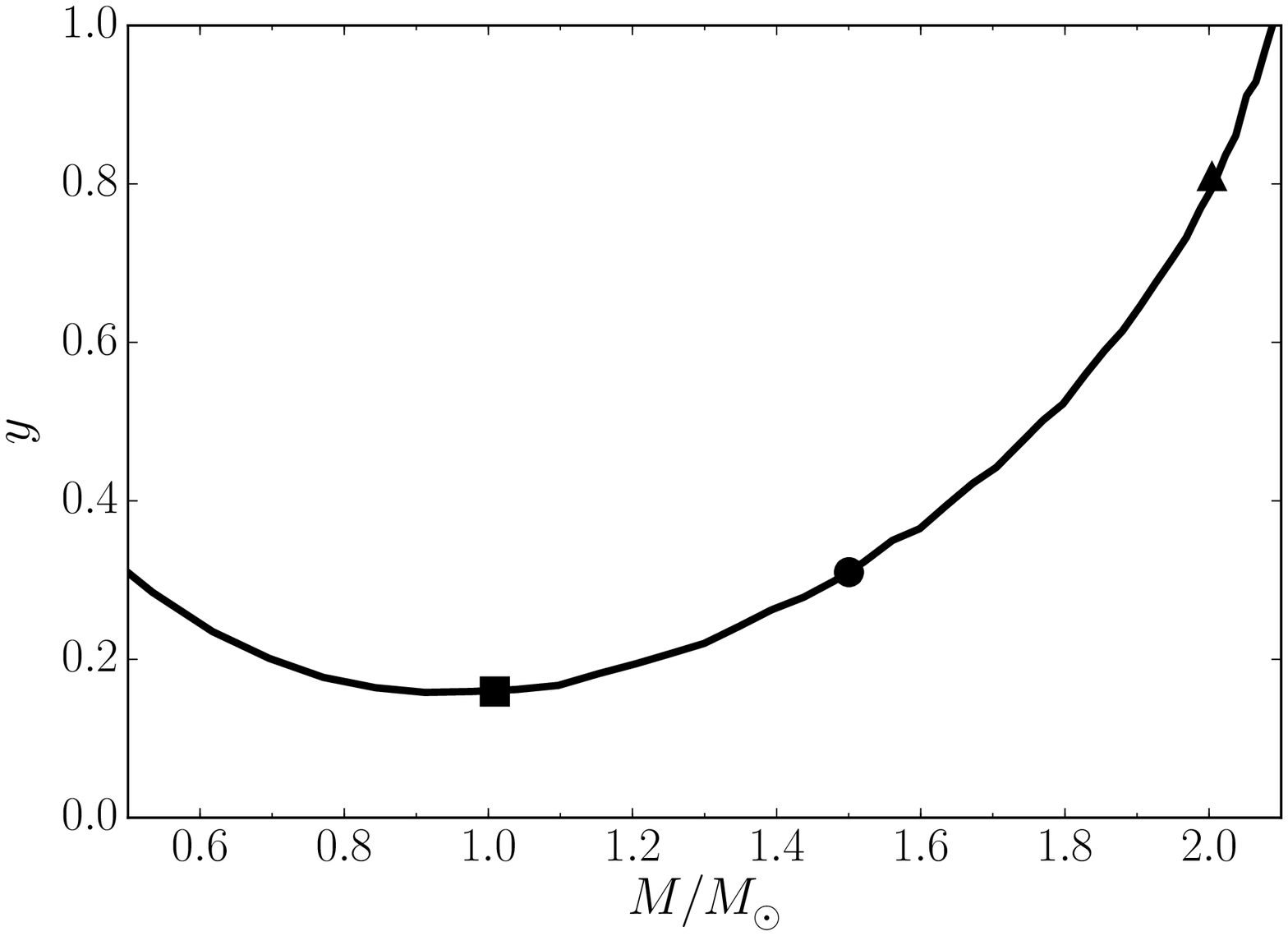}}
\caption{(a) Compactness vs. Mass. (b) $y$ vs. Mass.}
\label{figCy}
\end{figure}

\begin{figure}[!t]
\begin{center}
\includegraphics[width=2.25in]{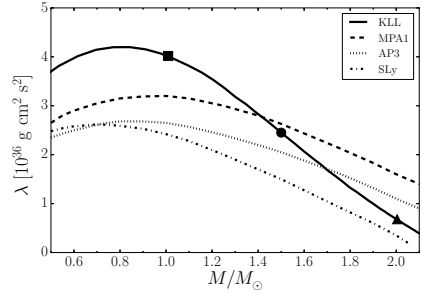}
\caption{ The tidal deformability parameter $\lambda$ in the mass range $0.5M_\odot - M_{max}$.}
\label{fig.7}
\end{center}
\end{figure}

By  solving TOV equation and Eq.\ (\ref{ODE}) together, we can then calculate  $y$ and the compactness $C$ (see Fig.\ \ref{figCy}) for the interior solution and  we  obtain  $\lambda$ or Love number, Eq.\ (\ref{lovek2}),   as shown in Fig.\ \ref{fig.7}.  The compactness $C=M/R$ in the range of mass $1.4$--$2 M_{\odot}$ is found to be 0.16 - 0.26 as shown in Fig.\ \ref{fig6-3}.

The deformability parameter  for $1.4M_{\odot}$ is found to be   $2.86 $. It can be compared with those of different EoS's  with only $npe\mu$ matter.   For example, the EoS's of SLy\cite{DH}, AP3\cite{AP3} and  MPA1\cite{MPA}  for the same mass give $\lambda = 1.70, ~2.22$ and $2.79$ respectively~\cite{PRD/81/123016}. On the other hand, the slope of $\lambda$ is found to be stiffer than those above in the mass range $1M_{\odot}$--$2 M_{\odot}$.    In the lower mass region around  $\lesssim 1M_{\odot}$, the deformability parameter is found to be relatively higher than those of above EoS's ($\lambda < 3$),  with the maximum value of $ 4.2$  at $0.84M_{\odot}$.  At higher mass region the deformability parameter in this work is lower than those EoS above mentioned.

It is interesting to note that recent numerical analysis\cite{read, HKSS} demonstrated the  measurability  of tidal deformations\footnote{In their analysis, the  dimensionless form of deformability parameter, $\Lambda = G (c^2 / GM)^5 \lambda$ has been used. In this work $\Lambda = 634$ for $M=1.35M_{\odot}$ .} determined by the change of late inspiral wave forms for $\delta \Lambda > 400$.

\section{Summary}
 We discussed the physical properties of stellar matter with a new stiffer EoS, which has been proposed recently using a new scaling law (new-BR/ BLPR) in medium caused by topology change at high density~\cite{Dong}, by  extending  Dong {\it et al.}'s work for pure neutron matter to a realistic nuclear matter of  $n$, $p$, $e$ and $\mu$.  The mass- radius and the tidal deformability were calculated.

The calculated maximum mass of compact star is found to be about $2 M_\odot$ with its radius about 11km.  The radius for the mass range of $1M_{\odot}$--$ 2 M_{\odot}$  is found to be $11.2$--$12.2 $km. The calculated deformability parameter for the stiffer EoS employed in this work is in the range  $  4.0$--$ 0.68$.

What characterizes the approach presented in this work is the stiffening of the EoS due to topology change predicted in the description of  baryonic matter with  skyrmions put on crystal background to access high density. The change is implemented in the properties of the parameters of the effective Lagrangian anchored on chiral symmetry and manifests in nuclear EFT formulated in terms of RG-implemented $V_{lowk}$. Given that the approach describes  fairly well the baryonic matter up to normal nuclear density, it is the changeover of skyrmions to half-skyrmions at a density $\sim$ (2--3)$n_0$ that is distinctive of the model used. This topology change involves no change of symmetries -- and hence no order parameters, therefore it does not belong to the conventional paradigm of phase transitions. But it impacts importantly on physical properties as described in various places in a way that is not present in standard nuclear physics approaches available in the literature. It is interesting to note that,  as has been discussed recently~\cite{hatsuda,baym}, there is another way to produce the stiffening in EoS to access the massive compact stars. It is to implement a smooth changeover from hadronic matter -- more or less well-described -- to strongly correlated quark matter, typically described in NJL model. By tuning the parameters of the quark model so as to produce a changeover at a density $\gsim 2 n_0$, it has been possible to reproduce the features compatible with the properties of observed massive stars.

After the detections of gravitational waves binary black holes, the  possible detection of  gravitational wave signals from coalescing binary neutron stars are well expected during the next run of aLIGO and aVirgo
and  the detection   will inform  us  of  the detailed effect of the tidal deformation\cite{PRD/77/021502,PRD/81/123016,read,HKSS}. Recently  the tidally modified waveforms have been  developed up to  the high frequency of merger\cite{read,bernucci}, such that the deformability parameter $\lambda$ , a function of the neutron-star EOS and mass, is measurable within the frequency range of the projected design sensitivity of aLIGO and aVirgo. It has been also demonstrated in Bayesian analysis that the tidal deformability can be measured to better than $\pm 1 \times 10^{36}$ g cm$^2$ s$^2$ when multiple inspiral events  from three detectors of aLIGO-aVirgo network~\cite{aLIGO,aVirgo} are analyzed\cite{Lackey}. They also show that the neutron star radius can be measured to better than $\pm 1$ km.  Thus  the simultaneous measurement of mass, radius and deformability using gravitational wave detectors could present an exciting possibility to eventually pin down the highly uncertain EoS for the nuclear matter in the mass range of $1M_{\odot}$--$2 M_{\odot}$. This would provide a probe for the state of baryonic matter at the high  density that is theoretically the most uncertain. And the hope is whether one can  confirm or falsify the strategies taken and assumptions made in \cite{Dong,LPR} and whether  the result would then help point the directions to be taken in the efforts described in \cite{Dong,Paeng}.

{\it Hyun Kyu Lee}: ``When Gerry invited me to Stony Brook  in 1998 for  my sabbatical year, he put me in a house just next to his. One late afternoon he came to  our place with a big smile and a basket of potatoes he just dug out in his yard.  He had keen interest in hearing the news of detecting gravitational waves, which was one of his favorite laboratories up in the sky.  I am now missing his big smile and a basket of comments on recent detections of gravitational waves, GW150914 and GW151226."


\section*{Acknowledgments}
The authors would like to thank  Mannque Rho, Tom Kuo and Won-Gi Paeng for helpful discussions and  acknowledge the hospitality at APCTP where a part of this work was done. The work was  supported in part by WCU project of Korean Ministry of Education, Science and Technology (Grant No. R33-2008-000-10087-0) and also in part by KISTI.



{}

\begin{thebibliography}{}

\bibitem{demorest} P.~B. Demorest {\it et al.}, Nature {\bf 467}, 1081 (2010)
\bibitem{antoniadis} J. Antoniadis {\it et al.}, Science {\bf 340}, 1233232 (2013)
\bibitem{PRL}B.~P.  Abbott et al. Phys. Rev. Lett. {\bf 116}, 061102; B.P. Abbott et al. Phys. Rev. Lett. {\bf 116}, 241103 (2016)
\bibitem{Dong} H. Dong {\it et al.}, Phys.\ Rev.\ C {\bf 87}, 054332 (2013)
\bibitem{LPR} H.~K. Lee, W.~-G. Paeng and M. Rho, Phys. Rev. D {\bf 92}, 125033 (2015)
\bibitem{Paeng} W.-G. Paeng, T.~T.~S.  Kuo, H.~K.  Lee and M.  Rho, Phys. Rev.  C {\bf 93}, 055203 (2016)
\bibitem{byp} H. ~K. Lee, B.~Y. Park and M. Rho, Phys. Rev. {\bf C 83}, 025206 (2011)
\bibitem{lr} H.~K.Lee and M.Rho, Int.\ J.\ Mod.\ Phys.\ E {\bf 22}, 1330005 (2013)
\bibitem{aligo} http://www.advancedligo.mit.edu
\bibitem{avirgo} http://www.cascina.virgo.infn.it/advirgo

\bibitem{rho} M. Rho, arXiv:1604.02662

\bibitem{BR} G.~E. Brown and M. Rho, Phys. Rev. Lett. {\bf 66}, 2720 (1991).
\bibitem{tom} H. Dong, T.~T.~S. Kuo, and R. Machleidt, Phys. Rev. C 80, 065803 (2009)
\bibitem{RhoG} H.~K. Lee and M. Rho, Eur. Phys. J. {\bf A 50}, 14(2014)
\bibitem{Astro.J/773/11} K.~Hebeler {\it et al.}, Astrophys.\ J. {\bf 773}, 11 (2013)
\bibitem{PhysRev/55/364} R.~C.~Tolman, Phys.\ Rev.\ {\bf 55}, 364 (1939)
\bibitem{PhysRev/55/374} J.~R.~Oppenheimer and G.~M.~Volkoff, Phys.\ Rev.\ {\bf 55}, 374 (1939)
\bibitem{PRD/81/123016} T.~Hinderer {\it et al.}, Phys.\ Rev.\ D {\bf 81}, 123016 (2010)
\bibitem{PRD/77/021502} E.~E.~Flanagan and T.~Hinderer, Phys.\ Rev.\ D {\bf 77}, 021502 (2008)
\bibitem{ApJ/677/1216} T.~Hinderer, Astrophys.\ J.\ {\bf 677}, 1216 (2008)
\bibitem{PRD/80/084035} T.~Damour and A.~Nagar, Phys.\ Rev.\ D {\bf 80}, 084035 (2009)
\bibitem{PRD/80/084018} T.~Binnington and E.~Poisson, Phys.\ Rev.\ D {\bf 80}, 084018 (2009)
\bibitem{read} J.~S. Read {\it et al.}, Phys. Rev. D {\bf 88}, 044042 (2013).
\bibitem{HKSS} K. Hotokezaka {\it et al.}, Phys. Rev. {\bf D93}, 064082 (2016).

\bibitem{PRD/58/124031} K.~S.Thorne, Phys.\ Rev.\ D {\bf 58}, 124031 (1998)
\bibitem{ApJ/149/591} K.~S.Thorne and A.~Campolattaro, Astrophys.\ J. {\bf 149}, 591 (1967)
\bibitem{PhysRev/108/1063} T.Regge and J.~A.Wheeler, Phys.\ Rev.\ {\bf 108}, 1063 (1957)

\bibitem{DH}F. Douchin and P. Haensel, A\&A 380, 151 (2001)
\bibitem{AP3}A. Akmal, V. R. Pandharipande, and D. G. Ravenhall, Phys. Rev. C {\bf 58}, 1804 (1998).
\bibitem{MPA}H. Muther, M. Prakash, and T. L. Ainsworth, Physics Letters B {\bf 199}, 469 (1987).

\bibitem{hatsuda}K.~Masuda, T.~Hatsuda and T.~Takatsuka,
  Astrophys.\ J.\  {\bf 764}, 12 (2013)
  [arXiv:1205.3621 [nucl-th]].
\bibitem{baym} T.~Kojo, P.~D.~Powell, Y.~Song and G.~Baym, Phys. Rev. D {\bf  91}, 045003 (2015)



\bibitem{bernucci} S. Bernuzzi, A. Nagar, T. Dietrich, and T. Damour, Phys. Rev. Lett. {\bf 114}, 161103 (2015)


\bibitem{aLIGO}G. M. Harry and LIGO Scientific Collaboration, Classical and Quantum Gravity 27, 084006 (2010).
\bibitem{aVirgo}F. Acernese, et al. (Virgo Collaboration),Advanced virgo baseline design, VIR-027A-09 (2009).
\bibitem{Lackey} B.~D. Lackey and L. Wade, Phys. Rev. D {\bf 91}, 043002 (2015)







\end{thebibliography}
\end{document}